\documentclass[pmlr]{jmlr}

\usepackage{longtable}

\usepackage{booktabs}
\usepackage[load-configurations=version-1]{siunitx} 

\newcommand{\real}{\ensuremath{\mathbb{R}}}
\newcommand{\ltwo}{\ensuremath{\mathbb{L}^2}}

\theorembodyfont{\upshape}
\theoremheaderfont{\scshape}
\theorempostheader{:}
\theoremsep{\newline}

\usepackage{natbib}

\jmlrworkshop{GeoMedIA}

\title[A WAE for White Matter Streamline Analysis]{StreamNet: A WAE for White Matter Streamline Analysis}

  \author{\Name{Andrew Lizarraga${}^{1,2}$}
  \Email{andrewlizarraga@g.ucla.edu}\hfill \\
  \Name{Katherine L. Narr${}^{1,3,5}$}
  \Email{narr@ucla.edu}\hfill \\
  \Name{Kirsten A. Donald${}^{4,7}$}
  \Email{kirsty.donald@uct.ac.za}\hfill \\
  \Name{Shantanu H. Joshi${}^{1,5,6}$}
  \Email{s.joshi@g.ucla.edu}\hfill \\
  \addr 1. Ahmanson Lovelace Brain Mapping Center, University of California, Los Angeles\\
  \addr 2. Department of Statistics, University of California, Los Angeles\\
  \addr 3. Department of Psychiatry and Behavioral Sciences, University of California Los Angeles\\
  \addr 4. Neuroscience Institute, University of Cape Town, Cape Town, South Africa\\
  \addr 5. Department of Neurology, University of California, Los Angeles\\
  \addr 6. Department of Bioengineering, University of California, Los Angeles\\
  \addr 7. Department of Paediatrics and Child Health, Red Cross War Memorial Children’s Hospital and University of Cape Town, Cape Town, South Africa\\
 }

\begin{document}

\maketitle

\begin{abstract}
We present StreamNet, an autoencoder architecture for the analysis of the highly heterogeneous geometry of large collections of white matter streamlines. This proposed framework takes advantage of geometry-preserving properties of the Wasserstein-1 metric in order to achieve direct encoding and reconstruction of entire bundles of streamlines. We show that the model not only accurately captures the distributive structures of streamlines in the population, but  is also able to achieve superior reconstruction performance between real and synthetic streamlines. Experimental  model performance is evaluated 
on white matter streamlines resulting from $T1$-weighted diffusion imaging of $40$ healthy controls using recent state of the art bundle comparison metric that measures fiber-shape  similarities.

\end{abstract}


\section{Introduction}
\label{sec:intro}

Recent technological advances in diffusion $T1$-weighted image acquisition and processing, as well as increasing number of population studies has resulted in availability of large new and diverse data sets. Consequently, there has been a been a recent rise in machine learning applications on these data sets. However, the analysis of white matter fiber bundles (collections of streamlines) still poses a unique challenge in that the data sets can be small, the data itself comprise high-dimensional objects that are not typically encountered in classical machine learning paradigms, especially if computations on  such streamline data are to preserve anatomical constraints. Notably, the individual streamlines themselves form highly heterogeneous shapes which are difficult to properly capture and encode in statistical models. This has necessitated the use of  geometric techniques in the analysis of these diverse data sets.


\subsection{Related Work}
There have been a variety of geometric techniques implemented to represent and analyze white matter streamlines. The approach by 
\cite{durrleman2011registration} proposes a current representation which allowed one to realize streamlines as differential forms on vector fields, which is then used to perform diffeomorphic registration between streamlines.  The approach by \cite{glozman2018framework} uses a simple point cloud representation of fiber bundles and analyzes their respective centroids. A recent approach by \cite{lizarraga2021alignment} represents streamlines as tangent spaces on square-root velocity manifolds to both preserve and compare geometry. Overall, these methods produce vastly different geometric representations of white matter streamlines and often must resort to computationally expensive data augmentation processes in order to represent white matter bundles in these frameworks. 

To reduce the data complexity and dimensionality, ideas such as implicitly representing geometric structure in low dimensional manifolds have been suggested. For example, high resolution reconstructions resulting from surface streamline data proposed by  \cite{Han2018-vq} makes use of autoencoders. But this approach requires training on domain specific data sets with small numbers of streamlines. Such approaches cannot be directly used for collections of highly varied streamlines like white matter due to the lack of consistency across subjects. Notably, \cite{LEGARRETA2021102126} suggested autoencoders for tract filtering, but could only make use of this technique on individual streamlines in order to meet the data demands required for training the network. Additionally the network is trained to output simple 1-dimensional arrays instead of preserving locality of fibers in bundles or cohesion of streamlines with respect to the entire bundle collection. A recent  approach by~\cite{Zhong2022} follows this work by implementing LSTM recurrent autoencoders to take advantage of the sequential structure of individual streamlines. This allows for a refined analysis of the latent space and further by performing cluster analysis from the latent space, one can extract the entire reconstructed bundle. However this approach is still trained on individual streamlines and ignores bundle cohesion. On the other hand, the approach by~\cite{Chandio2021.10.26.465991} represents white matter tracts by pairwise comparisons of streamlines  to form a distance matrix, and then use techniques such as t-SNE and UMAP to reduce the bundle in a low dimensional manifold to detect outlier streamlines. While this captures the structure of the white matter bundle, the pairwise computation of streamlines can be computationally expensive, especially for high-resolution bundles.

\subsection{Contributions}
In this paper, our goal is to not only encode geometric structures of streamlines, but also to preserve the spatial locality of the streamlines in a fiber bundle, which we term as the  bundle cohesiveness property of the collection. Instead of achieving an embedding of streamlines in a latent space, we want a direct encoding that allows one to sample and synthesize full bundles that not only preserve geometric structures, but also allow to reconstruct streamlines that respect the bundle cohesiveness property. This paper makes the following contributions: 1) we develop a deep neural network architecture, StreamNet, that directly encodes bundles of streamlines in a latent manifold. 2) the heterogeneous  geometry of the streamlines is implicitly preserved by making use of the Wasserstein-1 metric as a ground metric. 3) 
the  proposed architecture differs from previous approaches that attempt to encode individual streamlines, or even characteristic fibers from a collection of streamlines.  We follow a similar approach to~\cite{pmlr-v97-dukler19a}, which compares image spaces with the Wasserstein metric. However unlike, their approach, which uses a generative adversial network (GAN), we propose an  autoencoder architecture which provides an advantage in that we can use less training data and avoid stability issues. To our knowledge, this is the first such deep learning approach that can perform direct encoding and reconstruction of entire collections of streamlines at once.

\section{Methods}

\subsection{Representation of Streamlines}
\label{sec:streamlines}
We focus on the 3D coordinates of white matter streamlines, while ignoring the diffusion parameters such as fractional anisotropy that are measured along the tracts. 
We denote a bundle of streamlines $\mathcal{S}$ by a set $\{X_i\}, i = 1, \ldots, N$, such that $X_i \in \ltwo([0, 1], \real^3)$, where $N$ is the number of streamlines in the bundle and $\mathbb{L}^2$ denotes the Lebesgue square integrable functions. For implementation purposes, we discretize $X_i$ and with a slight abuse of notation, represent it to denote an array  $X_i \in \real^{3 \times 100}$, where $100$ is the number of discrete points on the streamline. Additionally, we uniformly sub-sample fiber bundles to produce tracts consisting of $300$ fibers each. Abusing notation, we denote this by $\mathcal{S} \in \mathbb{R}^{300\times 3 \times 100}$. These arrays are then normalized and bounded within the the unit cube $[0,1]^3$. With this representation, we assume the support of any particular group of streamlines is 0 outside the array of definition. This representation allows one to  view streamlines as a distribution $\mathbb{P}_{\mathcal{S}}$ which will be used in order to compare streamlines via the Wasserstein-1 loss during training of the model.

\subsection{Wasserstein-1 Metric}
There are a variety of Wasserstein Autoencoder architectures (WAE's), but this paper differs in a couple of
aspects from typical  WAE's that must be noted. The WAE originally presented in \citep{tolstikhin2017wasserstein} suggests an $L^2$ loss on the input-output pair with ELBO loss via the Wasserstein metric on the bottle-neck distributions.
However this paper follows a hybrid approach between the traditional AE and the Wasserstein of Wasserstein Loss (WWL) as implemented in \citep{pmlr-v97-dukler19a}. The WWL was demonstrated to be an appropriate model by treating the sample space of images as a discrete space of normalized pixels for effective image retrieval. Following suit, we make use of the Wasserstein metric on the image-output pairs, but don't make comparisons on the latent distributions. This is possible because our bundles are $300 \times 3 \times 100$ arrays which can then be broadcast to $300 \times 300$ discretized and normalized images. This lets us compute the Wasserstein loss on the images space as suggested in \citep{pmlr-v97-dukler19a}. The Wasserstein-1 loss ($W_1$) is given by the following \citep{NIPS2015_a9eb8122} formula:

\begin{equation}\label{eq:w1loss}
W_1(\mathbb{P}_{\mathcal{S}_1}, \mathbb{P}_{\mathcal{S}_2}) = \inf_{\gamma \in \prod(\mathbb{P}_{\mathcal{S}_1}, \mathbb{P}_{\mathcal{S}_2})} \int_{K \times K} l_1(k_1, k_2) d\gamma(k_1, k_2),
\end{equation}
where $\prod(\mathbb{P}_{\mathcal{S}_1}, \mathbb{P}_{\mathcal{S}_2})$ is the set of joint probability measures on $K \times K$ having $\mathbb{P}_{\mathcal{S}_1}$ and $\mathbb{P}_{\mathcal{S}_2}$ as marginals, and $l_1 : K \times K \rightarrow \mathbb{R}$ is the cost function given by the $L^1$ norm $l_1 = |k_1(x) - k_2(x)|$, where $x$ is in the support of $K$, and $K$ is a finite set possessing metric structure induce by $l_1$. This is typically viewed as a computationally difficult optimal transport problem, however, as suggested by \cite{feydy2019interpolating}, we may use the Sinkhorn Divergence to effectively sample the image distributions to approximate the Wasserstein loss efficiently. We use an implementation of this approximation in the GeomLoss package \cite{feydy2019interpolating}, and KeOps library \cite{JMLR:v22:20-275}.



\section{Model Design}

\subsection{StreamNet Architecture}
\label{sec:architecture}

The StreamNet architecture operates on a bundle of streamlines, which are represented by discretized arrays of dimension $N \times 3 \times T$. In this paper, we re-sample all bundles down to $N=300$ streamlines, and discretize the streamlines to be composed of $T=100$ points.
\figureref{fig:network} shows the schematic of the architecture. The network is composed of 2D-convolutional layers which have LeakyReLU activations with a slope of $0.01$. The bottleneck is composed of fully connected layers with LeakyReLU activations and slopes of $0.01$, where the latent vector is a $256$ dimensional. We note that by virtue of using a sampled Sinkhorn approximation of the Wasserstein-1 loss, we gain regularity since the network only sees a few samples from the output distribution as opposed to the entire image itself~\citep{feydy2019interpolating,cuturi2013sinkhorn,luise2018differential}.

\begin{figure}[htbp]
\floatconts
  {fig:network}
  {\caption{Schematic of StreamNet Architecture.}}
  {\includegraphics[width=0.9\linewidth]{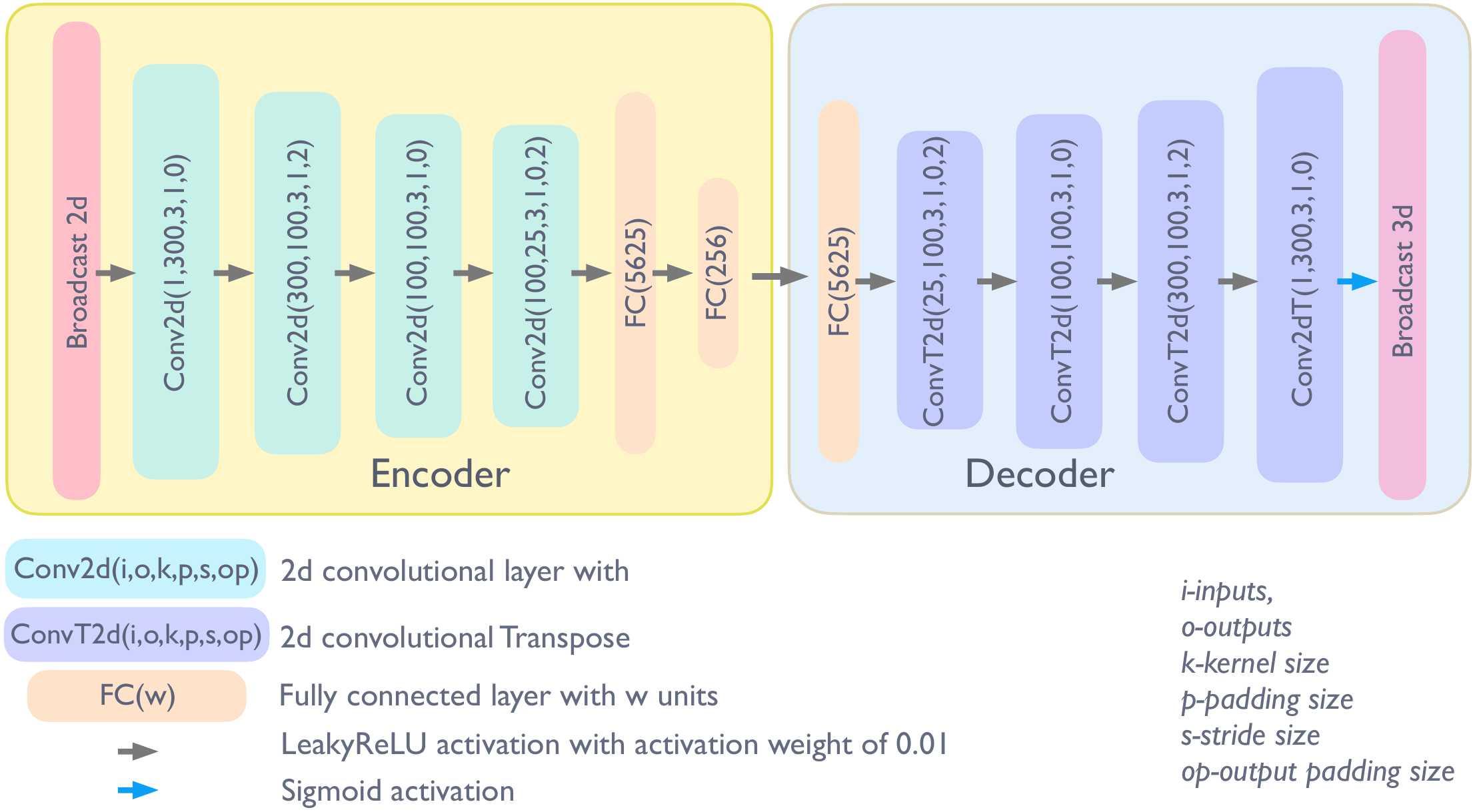}}
\end{figure}

\subsection{Training and Implementation Details}
All network weights are initialized via the Kaiming initialization \citep{7410480}. We only use batches of size $1$ as this is a limitation of sampling with Sinkhorn divergence \citep{feydy2019interpolating}. We used the Adam optimizer \citep{Kingma2015AdamAM} with a learning rate of $10^{-3}$ and a learning rate scheduler with a patience of 5 epochs and a decrement value of $0.2$. Additionally the network was trained on $4000$ training samples (explained in detail in \sectionref{sec:data}), with a validation set of size $400$ and a test set of size $400$. Finally, the network was trained for $200$ epochs. StreamNet was implemented in PyTorch \citep{paszke2019pytorch} on AMD Ryzen Threadripper 3960X 24-Core Processor @ 3.8 GHz machine with NVIDIA A100 GPU's. This model will be made publicly available on GitHub.

\section{Experimental Results}

\subsection{Data Collection and Processing}
\label{sec:data}
The data for all experiments was generated from diffusion-weighted images (DWI) by scanning $40$ healthy adult subjects (sex: 20M/20F) on a 3T Siemens PRISMA scanner with a 32-channel head
coil. The DWI protocol comprised a spin-echo echo planar sequence
(EPI), which included 14 reference images (b = 0 s/mm2), and multishell images
(b = 1500, 3000 s/mm2) with 92 gradient directions along with a T1-weighted image (voxel size=0.8mm3). All images were processed through  the Human Connectome Project minimal preprocessing pipeline \cite{GLASSER2013105}. Whole brain tractography was performed using multi-shell multi-tissue constrained spherical deconvolution followed by filtering of the tractograms. This resulted in each subject ranging between $6$-$10$ million fibers. The whole brain tractograms were then segmented via Automated Fiber Quantification (AFQ)~\citep{Yeatman2012TractPO} into 16 major groups (Left and Right except for the corpus callosum, respectively): Thalamic Radiation (Th Rad), Corticospinal Tract (CST), Cingulate Cingulum (CnCn), Corpus Callosum Forceps Major/Minor (CC F Maj/Min), Superior/Inferior Longitudinal Fasciulus (SLF/ILF), Uncinate (Unc), and Arcuate (Arc). This process also resulted in filtering  white matter fiber tracts for biologically relevant fibers.

A single tract can comprise of hundreds to thousands of streamlines. We uniformly sampled batches of $300$ streamlines from each of the $16$ tracts in order to create a testing data set consisting of equal sized streamlines from each tract type. 

We pruned the dataset to have an equal frequency of each tract to train the autoencoder on to prevent favoring the more dense tracts. For this given dataset this process resulted in each subject producing $75$ (or more) substracts per original tract, but we floored down to 75 tracts per subject to ensure equal frequency of the tracts when training. Thus there are $40 \times 16 \times 75 = 48,000$ tracts in total where each tract consists of 300 streamlines.
This reduces the data set to $4800$ sample bundles, where each sample point is an array of size $300 \times 3 \times 100$. This data set is split into a training set of $4000$ samples, a validation set of $400$ samples, and a test set of $400$ samples.
While this is a small training set compared to typical data sets used for training, the considerably large number of sample points across subjects helps to capture variability in the tracts.
This in turn influenced our decision to train the model for no more than $200$ epochs to reduce the chance of over-fitting. Given these constraints, the model obtained reasonable performance on the test set. Initially the network produced $W_1$ loss values in the range of 6-6.5, and by the end of training, the test set performance reported a $W_1$ loss of 0.5-0.6. It's difficult to intuitively interpret the drop in the $W_1$ loss, so we compare our results against state of the art streamline distance metrics and provide visual reconstructions of the original streamlines for qualitative comparisons.

\subsection{Streamline Reconstructions}
Here we demonstrate that that our model trained with the $W_1$ loss can produce anatomically relevant reconstructions of input bundles. Typically, the task of comparing similarities of bundles of streamlines is challenging,  and only a few metrics exist, however a new state of the art Bundle Analytic Score (BUAN) \cite{Chandio2020BundleAA} was developed specifically for this purpose. We present both quantitative findings of BUAN comparisons as well as visual quality of reconstructions on the test set. \figureref{fig:original_recon_img} shows an example of $9$ input bundles and their reconstructed outputs. Overall, we see that the model captures  the general structure of the streamlines and preserves the bundle cohesive property that each collection is composed of individual fibers. 

In \figureref{fig:buan_recon_chart} \textbf{A.} we see that the reconstructed tracts produce a BUAN score ranging from $0.9$ to $1.0$, with CCF Maj producing the majority of the outliers. By analyzing \figureref{fig:buan_recon_chart} \textbf{B.} we see an outlier Unc reconstruction (purple) misses some of the finer detail of the original tract. Missing large structures of a tract can result in a dramatically lower BUAN score despite the tracts having a visually similar overall shape. Tracts such as L-ILF, R-ILF, L-SLF, R-SLF, tend have a high variance in their fiber placement, resulting in larger variances of BUAN scores. The network was not trained with respect to BUAN scores, however given the sores are within $0.1$ of a perfect score $1.0$,  we suggest that the model is capturing and preserving the geometric structure of the full input streamlines despite only being given partial samples from the streamlines during training.
\begin{figure}[htbp]
\floatconts
  {fig:original_recon_img}
  {\caption{Original input (red) tracts from the test set compared to their reconstructed outputs (blue) from  StreamNet.}}
  {\includegraphics[width=0.9\linewidth]{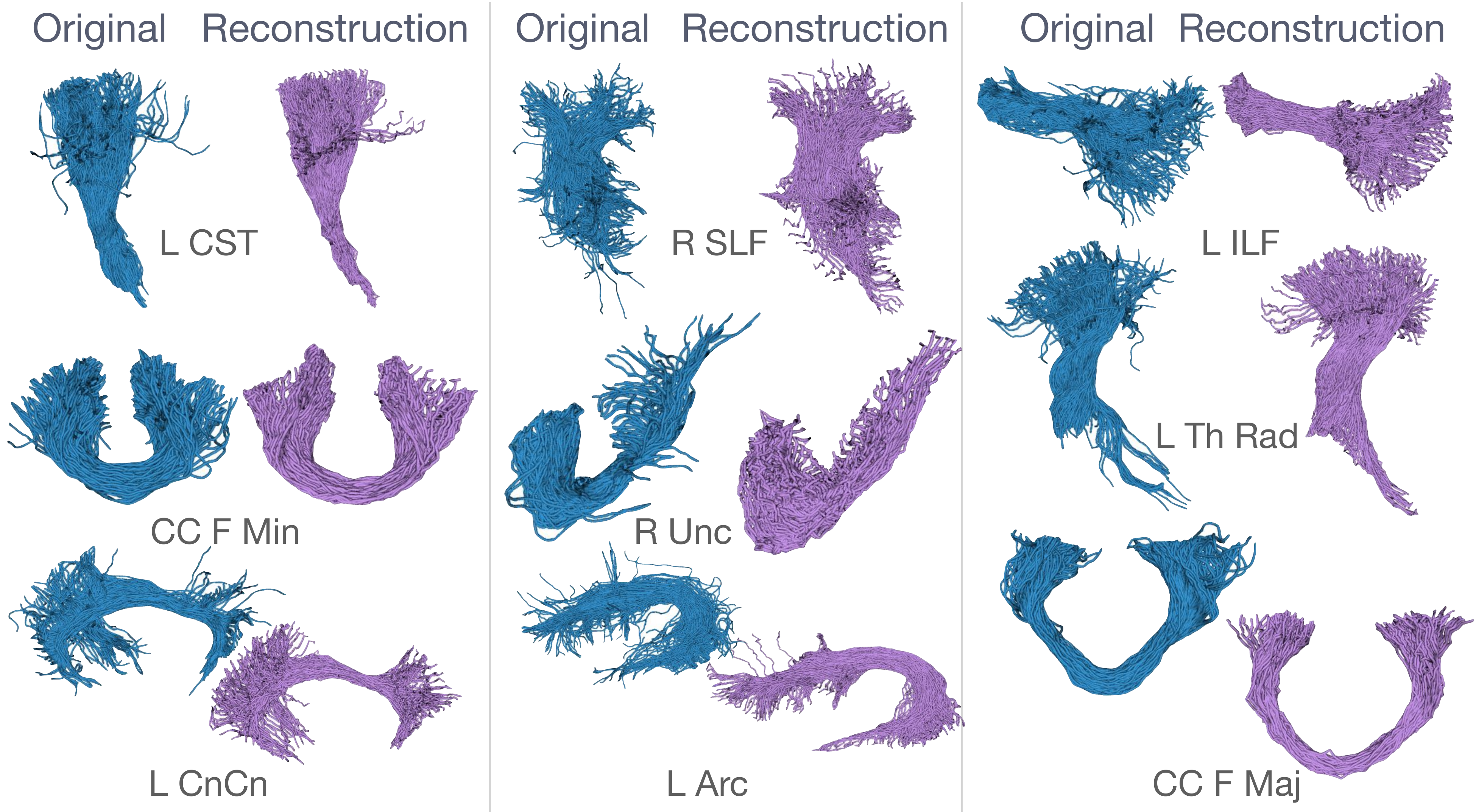}}
\end{figure}

\begin{figure}[htbp]
\floatconts
  {fig:buan_recon_chart}
  {\caption{\textbf{A. }Buan Scores comparing the original streamlines to the reconstructed streamlines, across all test samples (closer to 1.0 is better). \textbf{B.} An inspection of the Uncinate tract shows that  the network captures the overall shape, however there are subtle differences that are not always captured leading to outliers.}}
  {\includegraphics[width=0.9\linewidth]{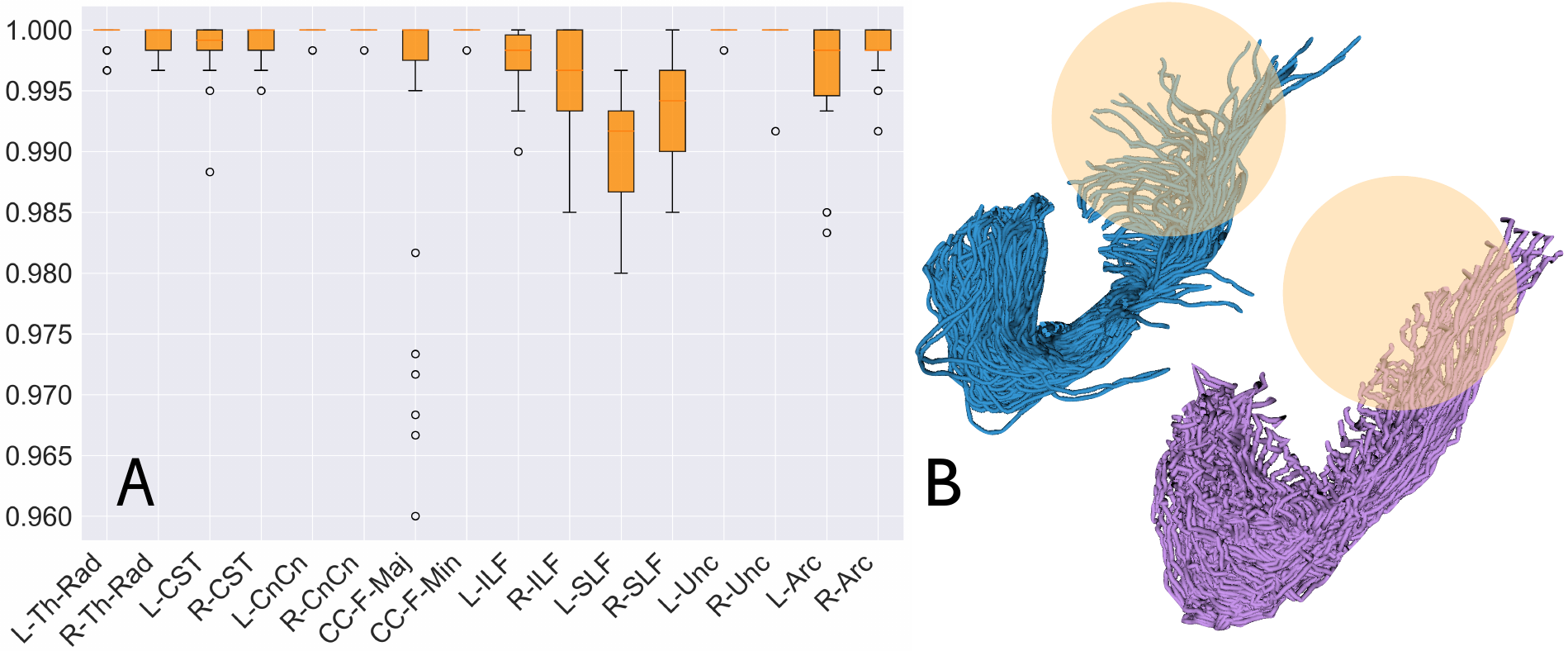}}
\end{figure}

\begin{figure}[htbp]
\floatconts
  {fig:latent_recons}
  {\caption{Bundles of streamline reconstructions from the latent averages compared to bundles from a random subject.}}
  {\includegraphics[width=0.9\linewidth]{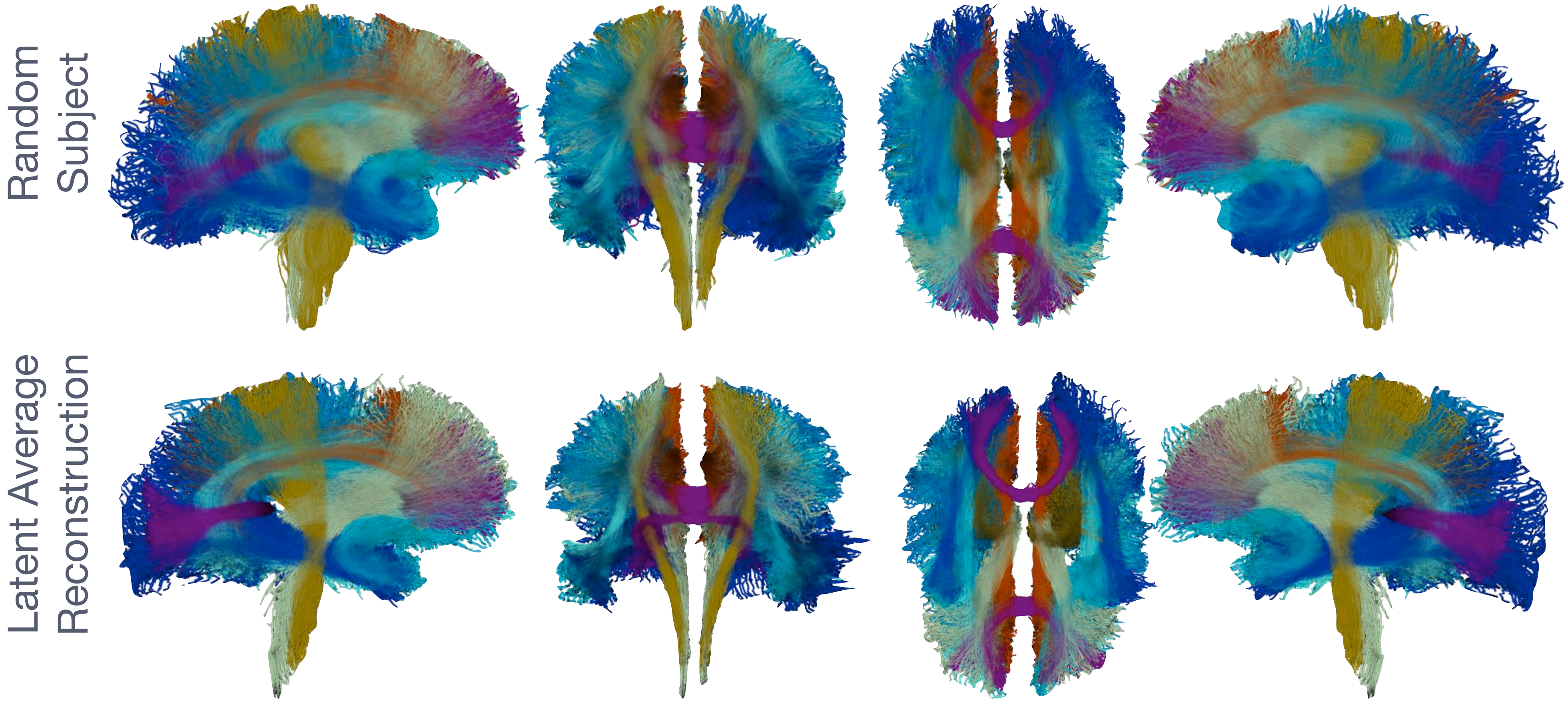}}
\end{figure}

\subsection{Bundle Templates in the Latent Space}
To better understand the model latent space, we compute averages of the resulting latent vectors from the encoded outputs of the original tracts per bundle. This results in $16$ such latent vectors which are then decoded by the model and are visualized and compared against the rest of the distribution. Upon visual inspection in \figureref{fig:latent_recons}, we see the decoded outputs of the average latent vectors and the corresponding subject bundles. The reconstructions from the average latent vectors appear anatomically plausible. Additionally, we compare the BUAN scores of average reconstructions against the original population, and project the latent vectors in $\mathbb{R}^2$ via t-SNE to better understand the separation of the data. As seen in \figureref{fig:latent_buan_average} \textbf{A.}, the latent projections are in separate groups. Note that the distance isn't necessarily preserved in the projection resulting from t-SNE, thus precluding any conclusions about the location of the embeddings (locations of L-CnCn and R-CnCn near each other for example). However the resulting latent space is vastly different from the latent space embeddings due to Legarreta et al.~\citep{LEGARRETA2021102126} and Zhong et al.~\citep{Zhong2022}. These works treat the sample points as individual streamlines and result in latent spaces that separate into individual groupings of fibers. Consequently, a bundle of fibers can be split up into multiple different latent vectors, so taking an average may not necessarily lead to a biologically valid reconstruction. Our model treats an entire bundles as a single sample point in the latent space, thus naturally leading to preserving bundle cohesiveness among streamlines and conveniently ensuring that average latent vectors for widely different tracts result in valid population template reconstructions for the correct tract type. Further, from  \figureref{fig:latent_buan_average} \textbf{B.}, we see that the BUAN scores are within the $0.88$  to $1.0$ range. We expect the BUAN scores to have a higher variance as we do not filter out outliers from the averages. 

\begin{figure}[htbp]
\floatconts
  {fig:latent_buan_average}
  {\caption{\textbf{A.} The embedding of latent-vectors in 2D. \textbf{B.} BUAN scores of latent averages compared to their respective sample populations.}}
  {\includegraphics[width=1\linewidth]{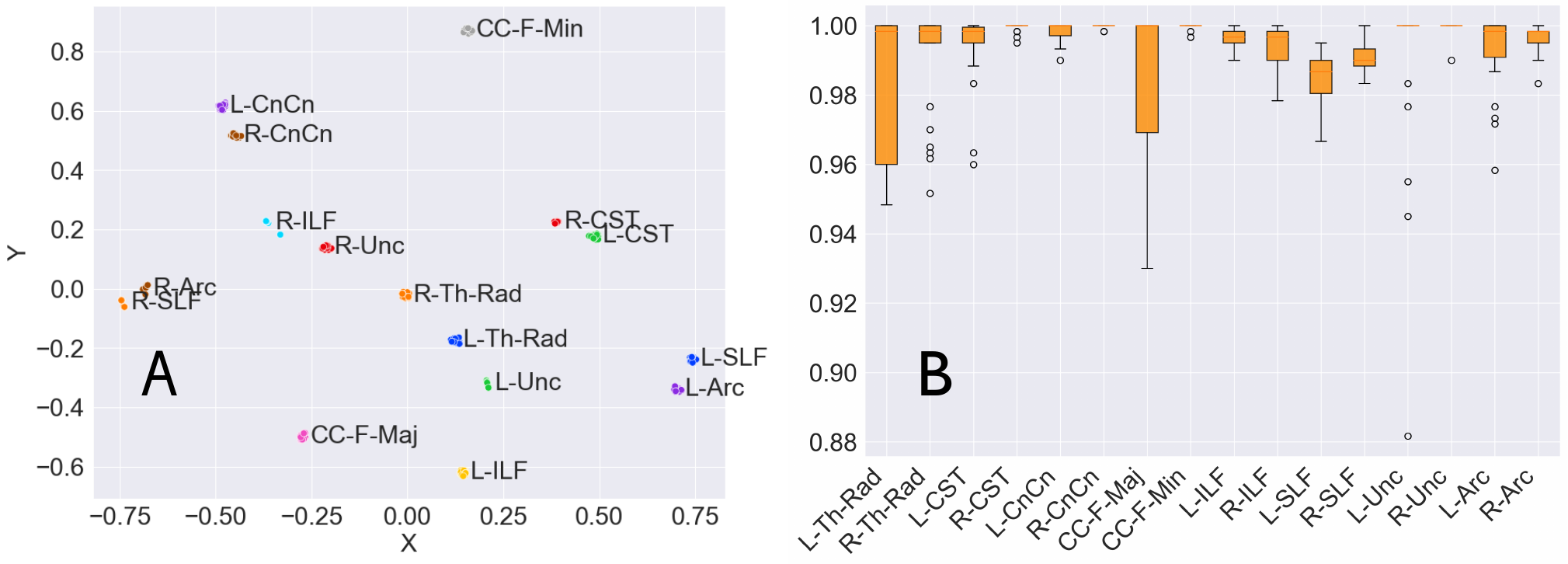}}
  
\end{figure}

\section{Discussion}
We demonstrated that a simple autoencoder architecture with $W_1$ loss can produce anatomically viable reconstructions for white matter streamlines. 
We validated these results by comparing reconstructions against the state of the art BUAN scores and visually see that the reconstructed results have overall similar structure. This also  suggests that the model is able to capture the heterogeneous geometry of the white matter streamlines accurately. Moreover, this proposed architecture allows one to construct population averages from samples within the latent space in order to produce white matter bundle  templates for future statistical inference. This is important for tractography modeling as the problem of bundle template estimation is  a challenging task that requires a pre-registration process across the entire population of white matter streamlines, which can quickly prove computationally intensive even for small data sets.

Given that the model is trained using an approximation of the $W_1$ metric via sampling with Sinkhorn divergence, the full input streamlines were not exposed to the model during training. Thus it's possible that some components of the bundle are never seen and the model interpolates under the constraints of the $W_1$ metric in order to preserve the fiber geometry. This may introduce variability in the reconstruction process as some fibers may be inferred or ignored by the model. However, a potential advantage is that the reconstructions appear smoother due to regularization provided by the Sinkhorn approximation. Thus the model can be also thought of a smoothing filter on the input data.

We envision that extensions of this simple model to more complex architectures such as Transformers or generative adversarial networks (GANs) can produce realistic sample populations from healthy control and treatment groups. This can in turn lead to potentially impactful applications for characterizing and understanding neurological manifestations of disease via tractoraphy among large population data sets.





\bibliography{pmlr-sample}


\end{document}